\documentclass[letters,usenatbib]{mnras}

\usepackage{newtxtext,newtxmath}
\usepackage{gensymb}

\usepackage[T1]{fontenc}

\DeclareRobustCommand{\VAN}[3]{#2}
\let\VANthebibliography\thebibliography
\def\thebibliography{\DeclareRobustCommand{\VAN}[3]{##3}\VANthebibliography}

\usepackage{graphicx}	
\usepackage{amsmath}	

\title[The H$\alpha$ sky in three dimensions]{The H$\alpha$ sky in three dimensions}

\author[L. McCallum et al.]{Lewis McCallum,$^{1}$
Kenneth Wood,$^{1}$
Robert A. Benjamin,$^{2}$
Dhanesh Krishnarao,$^{3}$
Catherine Zucker, $^{4}$
\newauthor
Gordian Edenhofer, $^{5}$
L. Matthew Haffner
 $^{6}$
\\
$^{1}$ School of Physics and Astronomy, University of St Andrews, North Haugh, St Andrews, KY16 9SS, UK\\
$^{2}$ Department of Physics, University of Wisconsin-Whitewater, Whitewater, WI 53190, USA\\
$^{3}$ Department of Physics, Colorado College, Colorado Springs, CO 80903, USA\\
$^{4}$ Center for Astrophysics, Harvard \& Smithsonian, 60 Garden St., Cambridge, MA 02138, USA \\
$^{5}$ Max Planck Institute for Astrophysics, Karl-Schwarzchild-Straße 1, 85748 Garching bei München, Germany \\
$^{6}$ Department of Physical Sciences, Embry-Riddle Aeronautical University, Daytona Beach, FL 32114, USA 
}

\date{Accepted XXX. Received YYY; in original form ZZZ}

\pubyear{\the\year{}}

\begin{document}
\label{firstpage}
\pagerange{\pageref{firstpage}--\pageref{lastpage}}
\maketitle

\begin{abstract}
We combine parallax distances to nearby O stars with parsec-scale resolution three-dimensional dust maps of the local region of the Milky Way (within 1.25~kpc of the Sun) to simulate the transfer of Lyman continuum photons through the interstellar medium. Assuming a fixed gas-to-dust ratio, we determine the density of ionized gas, electron temperature, and H$\alpha$ emissivity throughout the local Milky Way. There is good morphological agreement between the predicted and observed H$\alpha$ all-sky map of the Wisconsin H$\alpha$ Mapper. 
We find that our simulation underproduces the observed H$\alpha$ emission while overestimating the sizes of H{\sc ii} regions, and we discuss ways in which agreement between simulations and observations may be improved.
Of the total ionizing luminosity of $5.84 \times 10^{50}~{\rm photons~s^{-1}}$, 15\% is absorbed by dust, 64\% ionizes "classical'' H{\sc{ii}} regions, 11\% ionizes the diffuse warm ionized medium, and 10\% escapes the simulation volume. We find that 18\% of the high altitude ($|b| > 30\degree$) H$\alpha$ arises from dust-scattered rather than direct emission. These initial results provide an impressive validation of the three-dimensional dust maps and O-star parallaxes, opening a new frontier for studying the ionized ISM's structure and energetics in three dimensions.
\end{abstract}




\section{Introduction}

\begin{figure*}
\includegraphics[width=1.0\textwidth]{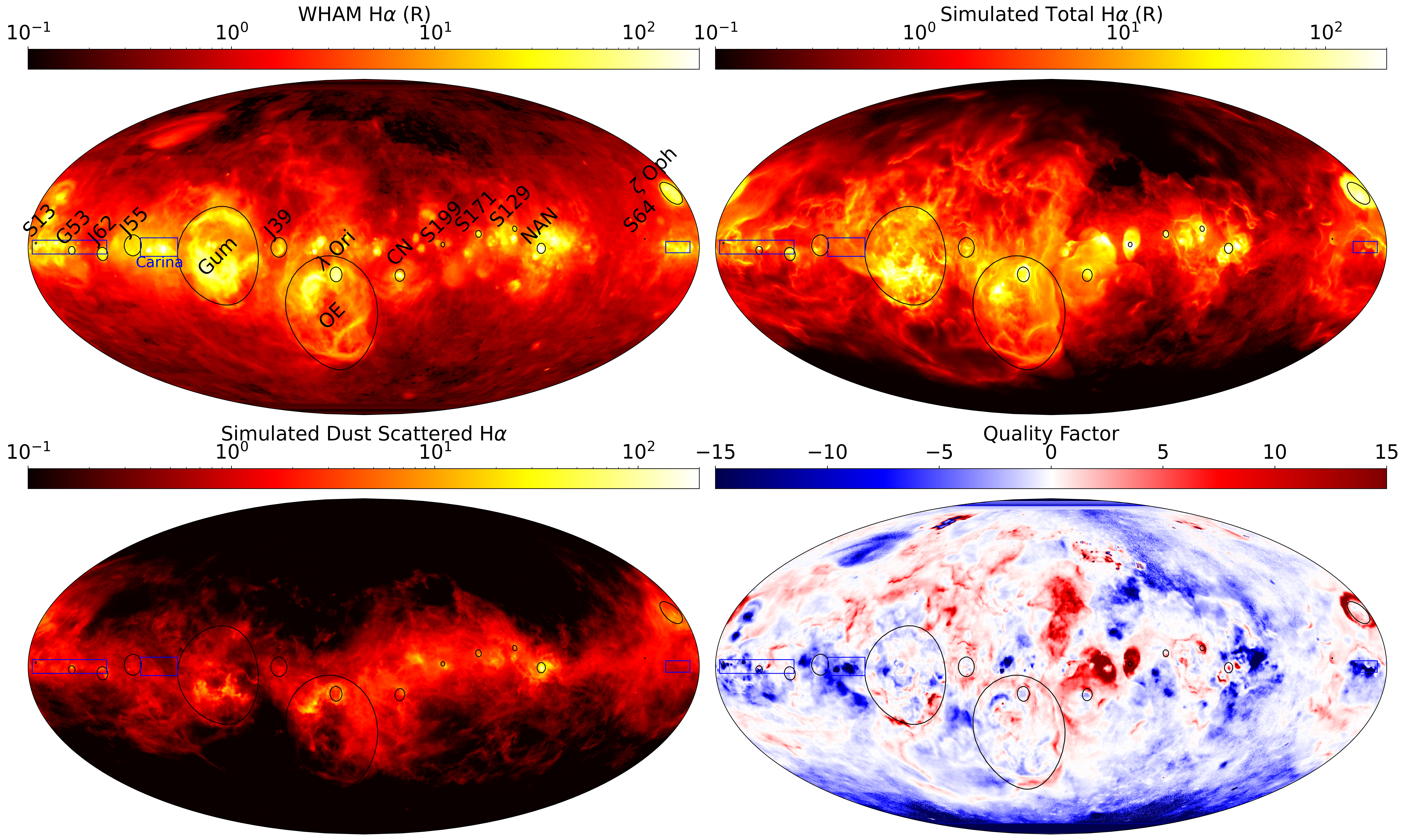}
\caption{Mollweide projections (centered on $l=180\degree$) of the the observed H$\alpha$ sky from WHAM (upper left), the simulated H$\alpha$ sky (upper right), the intensity of simulated H$\alpha$ which originates from dust scattering rather than direct H$\alpha$ emission (lower left), and the quality factor of simulated H$\alpha$ surface brightness as described in section~\ref{qf} (lower right). Individual regions are circled in the H$\alpha$ maps by their numbers in the catalogues by Gum (G), Sharpless (S), and Jardine (J)  with the exception of the following named regions; the Orion-Eridanus superbubble (OE), the Gum nebula (Gum), the California nebula (CN), the North American Nebula (NAN), the $\zeta$ Ophiuchi H{\sc ii} region ($\zeta$ Oph) and the $\lambda$ Orionis H{\sc ii} region ($\lambda$ Ori). Blue rectangles on the upper panels highlight regions of high H$\alpha$ in the WHAM map originating from outside our 1.25~kpc radius simulated volume, and thus do not appear in our simulated sky.
\label{fig:halphasky}}
\end{figure*}

Understanding the creation and transport of Lyman continuum photons from massive stars is a key element in determining the nature of galaxies as star formation engines. This ionizing radiation is mostly reprocessed into hydrogen and helium recombination and thermal dust emission, while some fraction of it can escape galaxies to produce the intergalactic radiation field. Measurement of the effects of this radiation allows us to quantify the star formation rate of galaxies and then compare to the structural and evolutionary factors that may regulate star formation \citep[see, for example][]{2012ARA&A..50..531K}.

Observations of the local Milky Way provide an opportunity to study this process in detail because we can resolve individual ionizing sources and the structure of the surrounding gas and dust. In the 1950s, a primary focus of these investigations lay in determining the spiral structure of the Milky Way by using wide-area H$\alpha$ sky surveys \citep{1952ApJ...115...89S,1955MmRAS..67..155G,1959ApJS....4..257S,1960MNRAS.121..103R} to identify nearby H{\sc ii} regions and measuring the distances to the exciting stars \citep{1953ApJ...118..318M}.

The original idealization of these H{\sc ii} regions as ``Stromgren spheres'' \citep{stromgren39} in a uniform medium has been modified over the subsequent decades as our understanding of the complexity of the interstellar medium increased. Massive stars are observed to form as part of clusters in dense molecular gas, but the combination of stellar winds, ionizing radiation and supernovae leads to a complex density structure for the interstellar medium, potentially filling the volume with networks of hot ($T \sim 10^{6}$~K) bubbles and superbubbles, interstellar chimneys, and ``galactic fountain'' flows \citep{1974ApJ...189L.105C,1976ApJ...205..762S,1977ApJ...218..148M,1980ApJ...236..577B,1989ApJ...345..372N}. 

This framework had to be modified when pulsars in globular clusters were used to demonstrate a significant mass of warm ionized gas extending a few kiloparsecs above the Galactic plane \citep{reynolds89}.\footnote{The existence of this gas had been posited to explain thermal absorption of low-frequency radio emission \citep{hoyle63}, but the height of this gas was observationally unconstrained.} This layer of Warm Ionized Medium (WIM), sometimes called the ``Reynolds layer'', or the diffuse ionized gas (DIG), exists in the volume beyond the classical H{\sc ii} regions, and analogous layers were detected in other edge-on spiral galaxies \citep{1990A&A...232L..15D,1990ApJ...352L...1R}.  In a series of papers, Reynolds outlined several mysteries associated with this gas: what powers it, what ionizes it, what supports it at heights greater than a thermal scale height for $10^{4}$~K gas, what fraction of ionizing photons emitted escape the galaxy, and so on \citep[see][]{haffner09}. These mysteries, combined with the realization that WIM emission was a significant source of foreground emission for cosmological studies \citep{reynolds92}, led to the construction of the Wisconsin H$\alpha$ Mapper (WHAM), which obtained the first velocity-resolved all-sky map of optical line emission from the diffuse ionized gas of the Milky Way \citep{haffner03}.

Although numerous investigations have addressed the issues above, the unknown values of the filling factors and density structure of different phases have limited our ability to make conclusive statements about the energization and physics of the ISM. Previous modeling efforts of LyC transport \citep{miller93,1994ApJ...430..222D,wood99,1999ApJ...510L..33B,2000ApJ...531..846D} explored the effects of different density structures, while \citet{zurita02} used H{\sc I} maps of NGC~157 as an input to explore the role of H{\sc ii} region LyC leakage in ionizing the WIM. However, recent development of maps of the three-dimensional distribution of dust in the local ISM \citep{2014ApJ...789...15S,2015ApJ...810...25G,2019ApJ...887...93G,2019A&A...625A.135L,2022A&A...661A.147L,2022A&A...664A.174V,2020A&A...639A.138L,edenhofer23} informed by parallax \citep{2016A&A...595A...1G,gaiadr3} and extinction \citep{zhang23} measurements for millions of stars  make it possible to study the transport of Lyman continuum (and other) photons, turning the local ISM into a laboratory for the physical processes regulating the structure and dynamics of the Galactic ISM.  

A realistic model of Lyman continuum transport would provide a 3D grid of free electron density and hence local contributions toward dispersion measure along any sight line, crucial in interpreting dispersions from extragalactic fast radio bursts. The free electron structure is also required for interpretation of surveys of Faraday rotation measure in constraining the 3D structure of the local magnetic field. Our simulations also estimate interstellar contributions of photoionized species with commonly used absorption lines (Mg, Ca, Na etc). They will help constrain Lyman continuum escape fractions with relevance to the study of cosmic reionization, give realistic fractions of ionizing photons lost due to dust absorption, and also help to identify the extent to which ionizing flux is able to escape individual H{\sc ii} regions. An estimate of direct versus scattered emission on every line of sight is also possible, as well as the ability to study the morphology and nature of individual regions of interest. It is also possible to determine contributions from individual ionizing sources, and constrain zones of influence of individual stars. Generating synthetic images also opens opportunities to fit for ISM and stellar parameters such as dust-to-gas ratio, ionizing luminosity and distances to every O star in the local volume.

In \S \ref{sec:simulation} we describe our simulations; in \S \ref{sec:results} we compare the resulting simulated H$\alpha$ sky maps to observations and derive some key, albeit preliminary, properties of the WIM, and in \S \ref{sec:future} we discuss numerous avenues for future investigations.

\section{Simulation Setup} \label{sec:simulation}

\subsection{Monte Carlo photoionization and H$\alpha$ dust scattering}

We use the Monte Carlo photoionization code \texttt{CMacIonize} \citep{cmacionize,cmi2} which is based on the earlier version of \citet{wood04}. \texttt{CMacIonize} simulates the transport of photons in a 3D density grid and calculates the equilibrium ionization and temperature structure for a given set of ionizing sources. We have updated the radiative and dielectronic recombination rates and include collisional ionization rates from the CHIANTI database \citep{delzanna21}. For each of the 10 iterations to convergence in the photoionization simulation, we utilize $10^{9}$ Monte Carlo photon packets.

Having computed the 3D temperature and ionization structure, we calculate the $\rm{H}\alpha$ emissivity using the tables of \citet{osterbock06}. The resulting H$\alpha$ emissivity grid is the input for a Monte Carlo simulation of dust scattering and absorption using a modified version of the code described in \citet{wood99}, with dust properties at H$\alpha$ from \citet{draine23}. The opacity of the dust grains at H$\alpha$ per hydrogen nucleon is set as $2.32\times10^{-22}~\rm{cm}^{2}~\rm{H}^{-1}$, the dust albedo was 0.72, and the Henyey-Greenstein factor for forward scattering was 0.44. The dust to gas ratio for both the photoionization simulation and dust scattering simulation was set as 0.00605 \citep{draine23}. For the dust scattering simulation we use $4\times10^{9}$ Monte Carlo photon packets.

\subsection{Density structure}

The 3D density structure was derived from the differential extinction maps within 1.25~kpc of the Sun \citep{edenhofer23}. This work releases twelve realizations of their 3D dust map, alongside the mean and standard deviation of the samples. Here we utilize the mean map. This work thus does not take into account the uncertainties in the dust map.

The maps are converted to total hydrogen density using the same technique as \citet{oniell24} (based on the the method \citet{zucker21}). We first use the extinction curves of \citet{zhang23} to translate the \citet{edenhofer23} map (in total extinction per parsec - $\rm{E}/\rm{pc}$) into \emph{Gaia} G-band extinction per parsec ($A_{G}/\rm{pc}$) via multiplication by a factor of 2.04. Then using the extinction to column density factor of $A_{G}/N_{H} = 4\times10^{-22}$ from \citet{draine09} gives $N_{H}/\rm{pc}$ of $5.1\times10^{21}~\rm{cm}^{-2}pc^{-1} \times E$, equivalently $1652~ \rm{cm}^{-3} \times E$. The resulting density was interpolated onto a regular $1024^3$ Cartesian grid using the interpolation script shipped with the data from \citet{edenhofer23}. This script interpolates the log-density structure and returns the exponential of the result.

The simulation box is a cube of side 2.5~kpc with the Sun at the centre.

\subsection{Ionizing sources}

Our photoionization simulation has 87 known O stars with good spectral classifications within 1.25~kpc of the Sun. We used the Galactic O-Star Catalog \citet{apellaniz13}\footnote{https://gosc.cab.inta-csic.es/galactic-o-star-catalog} and an additional 10 O stars from a compilation of H{\sc ii} regions and associated ionizing sources by K. Jardine\footnote{http://galaxymap.org/drupal/}. The Wolf-Rayet star contained within $\gamma^{2}$ Velorum in the Gum Nebula was also included. Barring nearby sources that have been spectrally mis-classified or deeply embedded sources, we anticipate that we have a reasonably complete sample of WR and O stars within 1.25 kpc.

Distances were obtained from Gaia parallaxes \citep{2020yCat.1350....0G} and earlier catalogs \citep{2002A&A...384..180F,2007A&A...474..653V,2012yCat.1322....0Z}. Of sources within 1.25 kpc, the mean statistical distance uncertainty was 9\% with only 9 sources with statistical uncertainties exceeding 20\%.
We discuss the effects of distance uncertainties on our results in \S \ref{sec:future}. 
 
The ionizing luminosities and effective temperatures of each source were determined by interpolating Tables 4-6 of \citet{martins05}. 
 
The effective temperatures were used to select an appropriate spectrum for each ionizing source from the WMBasic library \citep{pauldrach01}. The ionizing luminosity and spectral shape of the WR star was taken from \citet{crowther07}. The effects of any missing O stars, B stars, and other sources are discussed in \S \ref{sec:future}.
Uncertainties in ionizing luminosities would change both the size of ionized regions and the total H$\alpha$ intensity generated. Variations in spectral type will have less effect on the simulated H$\alpha$ sky, but might alter the structure of other ionized species which are not analyzed in this preliminary work.

\section{Results} \label{sec:results}

\subsection{The $\rm{H}\alpha$ Sky: The Inside and Outside View}
\label{qf}
Figure~\ref{fig:halphasky} shows the simulated all-sky map of $\rm{H}\alpha$ from ionized gas within 1.25 kpc, alongside the total $\rm{H}\alpha$ emission measured by WHAM \citep{haffner03,haffner10}. These survey data are accessible via \texttt{WHAMPY} \citep{whampy}. We have applied a velocity cut to the WHAM data at $\pm 12 ~\rm km ~s^{-1}$ to eliminate emission from outside the simulated volume. This is a helpful first order approximation, but a full knowledge of the thermal and kinetic components in the regions near the boundary of the region is required to completely remove spatially non-local emission. Also shown is the intensity of H$\alpha$ arising from dust scattering. The H$\alpha$ excess/deficit is also displayed using a \emph{quality factor}, $Q$,  defined as,
\begin{equation}
    \begin{cases} 
Q = \frac{I_{\rm{SIM}}}{I_{\rm{OBS}}} - 1 & I_{\rm{SIM}} > I_{\rm{OBS}},\\
Q = -(\frac{I_{\rm{OBS}}}{I_{\rm{SIM}}} -1) & I_{\rm{SIM}} < I_{\rm{OBS}}
\end{cases}
\end{equation}
for simulated intensity $I_{\rm{SIM}}$ and WHAM observed intensity $I_{\rm{OBS}}$. This is such that $Q = 0$ represents a perfect match to the observational data, positive values represent a multiplicative factor of overestimation of H$\alpha$, and negative values represent a multiplicative factor of underestimation. Several regions are annotated in the WHAM image.

\begin{figure}
    \centering
    \includegraphics[width=\columnwidth]{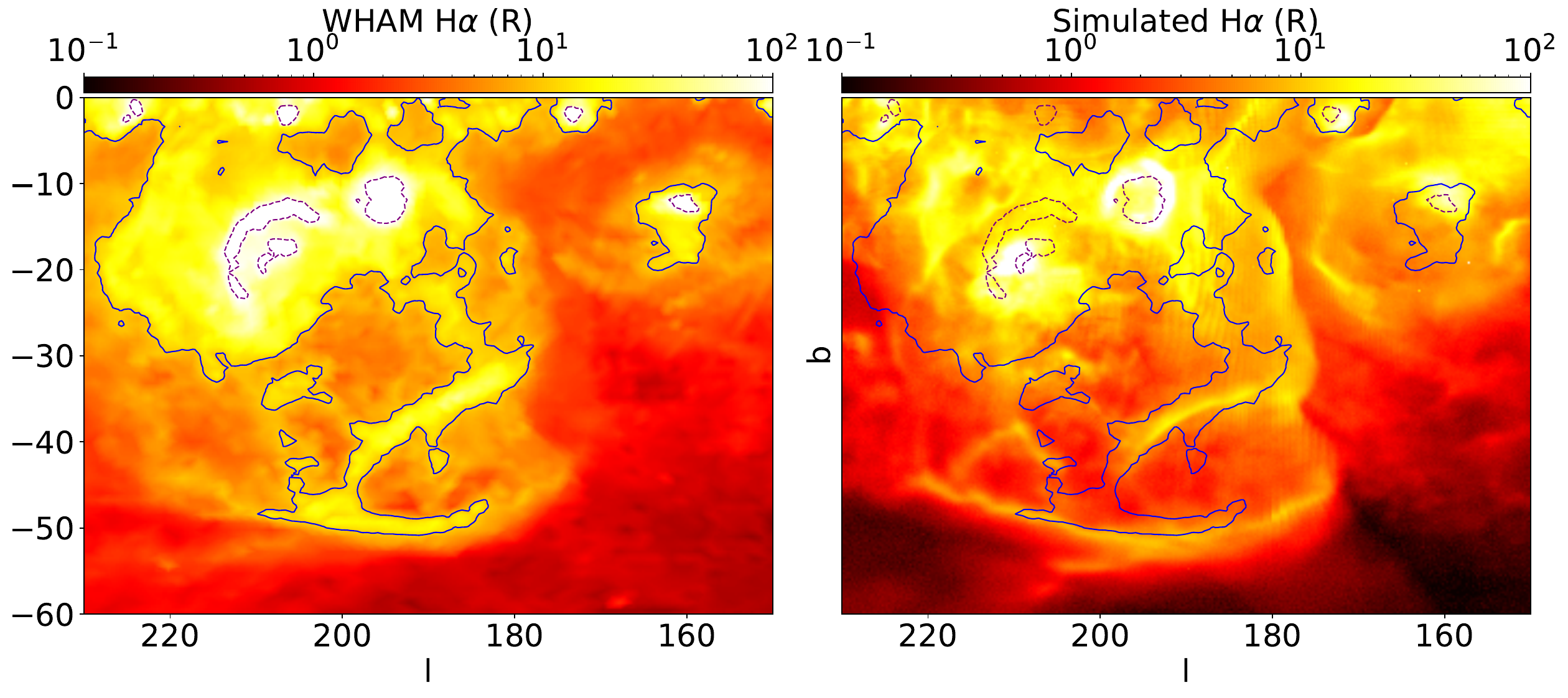}
    \caption{Zoom-in of the Orion-Eridanus superbubble in H$\alpha$ from the WHAM sky (left) and our simulated sky (right). The solid blue contours on both images show lines of 10 R brightness in the WHAM sky, and purple dashed contours are 100 R on the WHAM image. The H$\alpha$ emission in the simulated view of Orion-Eridanus is slightly more extended than observed. }
    \label{fig:oebubble}
\end{figure}

There is a good morphological agreement between the observations and simulations. Of the 87 O stars, 65 can be associated with 45 H$\alpha$-selected H{\sc ii} regions from WHAM, 38 of which are are reproduced in our simulations, including the $\zeta$ Oph (Sh 2-27) and $\lambda$ Ori (Sh 2-264) H{\sc ii} regions, the California (Sh 2-220) and North American Nebula (NGC 7000/Sh 2-117), the Orion-Eridanus superbubble and the Gum nebula. A zoom-in image of the Orion-Eridanus superbubble region (Fig \ref{fig:oebubble}) shows good agreement, reproducing several arcs and shells, although Barnards Loop is less prominent in the simulations than the observations. Of the seven nearby H{\sc ii} regions that are not reproduced by the simulation, most are in the inner Galaxy, $|l|<30^{\circ}$; the dust maps and O-star distances may merit re-examination in this direction. 

Most of the features in the WHAM map that do not appear in the simulations are for understandable reasons. Several high latitude H{\sc ii} regions are ionized by B stars, while other regions lie beyond the 1.25~kpc distance range of the 3D dust map, notably the $l=10-30^{\circ}$ Sagittarius ``spur'' \citep{2021A&A...651L..10K} and the $l=280-300^{\circ}$ Carina Arm ``tangency'' region. Because the emission lines from these regions are bright and broad, they contaminate low-velocity local emission. Gaussian fitting is required to remove this distant emission.

The H$\alpha$ excess/deficit shows that over most of the sky the simulation under-predicts the observed H$\alpha$ emission, typically by a factor of two. There are three principal reasons why this might be the case.  First, we only consider ionization by local O stars. Including emission from B stars, subdwarf OB stars, hot white dwarfs and other sources of LyC radiation will increase the predicted H$\alpha$ emissivity. Second, especially at lower latitudes, our simulations are missing H$\alpha$ emission from beyond 1.25 kpc. Similarly, any photons which would enter the box from outside the simulation volume are not included, so the H$\alpha$ in the box would be expected to be in deficit, in particular towards the edges and top/bottom of the simulation box.

Some regions of very low H$\alpha$ surface brightness are present in the simulated sky. In a future study, these low emission measure sight-lines will be compared with the free-electron column density from pulsar dispersion measures.

Given the good morphological agreement of the 3D simulation with the H$\alpha$ sky-view, one can create reliable views of the H$\alpha$ emission from different vantage points. Figure~\ref{fig:faceon} provides the face-on H$\alpha$ view of our location in the Milky Way as seen by an external observer. Some H{\sc ii} regions are compact and in dense and dusty regions; others are larger and more ring-like, including two prominent examples of broken rings near the Sun: the Orion-Eridanus superbubble and the Gum Nebula that appear to be venting into Galactic supershell GSH 238+00+09. The face-on view also shows regions of diffuse ionization, i.e. the Warm Ionized Medium. These images are reminiscent of views of H~{\sc{ii}} regions in nearby galaxies, e.g. \citet{1983AJ.....88..296H}, \citet{1998ApJ...506..135G} or \citet{2023MNRAS.520.4902G}.\footnote{For a particularly striking comparison, compare Figure~\ref{fig:faceon} to this amateur astronomer panorama of the LMC: \url{https://www.cielaustral.com/galerie/photo95.htm}} 

\subsection{The $\rm{H}\alpha$ Sky: Individual Regions}
\label{individualregions}

Although the simulated all-sky map shows good agreement with WHAM, an examination of individual H{\sc ii} regions points the way to future improvements. We highlight five individual regions that have been extensively studied, refer readers to their respective entries in the two-volume Star Formation Handbook \citep{2008hsf1.book.....R,2008hsf2.book.....R} for more information.

{\it $\zeta$ Ophiuchi H{\sc ii} region (Sh 2-27)}: This is the nearest O-star ionized H{\sc ii} region (at a distance of 135~pc), with slightly-elliptical $R=5^{\circ}$ bright H$\alpha$ emission centered at $l=5.8^\circ, b=+23.7^\circ$. $\zeta$ Oph is a ``runaway'' O star associated with Sco OB2 \citep{2008hsf2.book..235P}. From the 3D grids of emission measure we find it is ionization bounded (all the ionizing photons are absorbed) and no photons escape to ionize the low-density diffuse ionized gas in agreement with prior modelling by \citet{wood05}. 

The simulated H{\sc ii} region has a lower intensity H$\alpha$ halo extending $3^\circ$ (7 pc) beyond the observed radius, but with similar ellipticity. This larger extent may be due to (1) too high an ionizing luminosity from the interpolated tables, (2) under-resolving higher density substructure in the nebula (as noted in \citet{zucker21}) or (3) a drop in the mean dust-to-gas ratio around $\zeta$ Oph, resulting in an underestimate of the gas density. Potentially in support of the final option, we note that the 3D dustmaps of \citet{edenhofer23} show a cavity of dust density centered on the expected location of $\zeta$ Oph, however this could also be a physical cavity due to the dynamical effect of ionizing photons from $\zeta$ Oph. Uncertainties in the distance to $\zeta$ Oph could also contribute. 

{\it Orion-Eridanus superbubble}: This large structure at $l=190^\circ,~b=-40^\circ$ \citep{2008hsf1.book..459B} is morphologically very similar to observations. The shape of the Eridanus loop at the negative latitude end of the bubble closely resembles the WHAM map, suggesting the 3D structure of the superbubble in the dust map is reliable. A previous 3D analysis focused on Barnards Loop \citep{2023ApJ...947...66F}, but without radiative transfer. Our simulations do not convincingly reproduce this structure, possibly as a result of uncertainties in the stellar distances. 

\begin{figure*}
    \centering
    \includegraphics[width=0.9\textwidth]{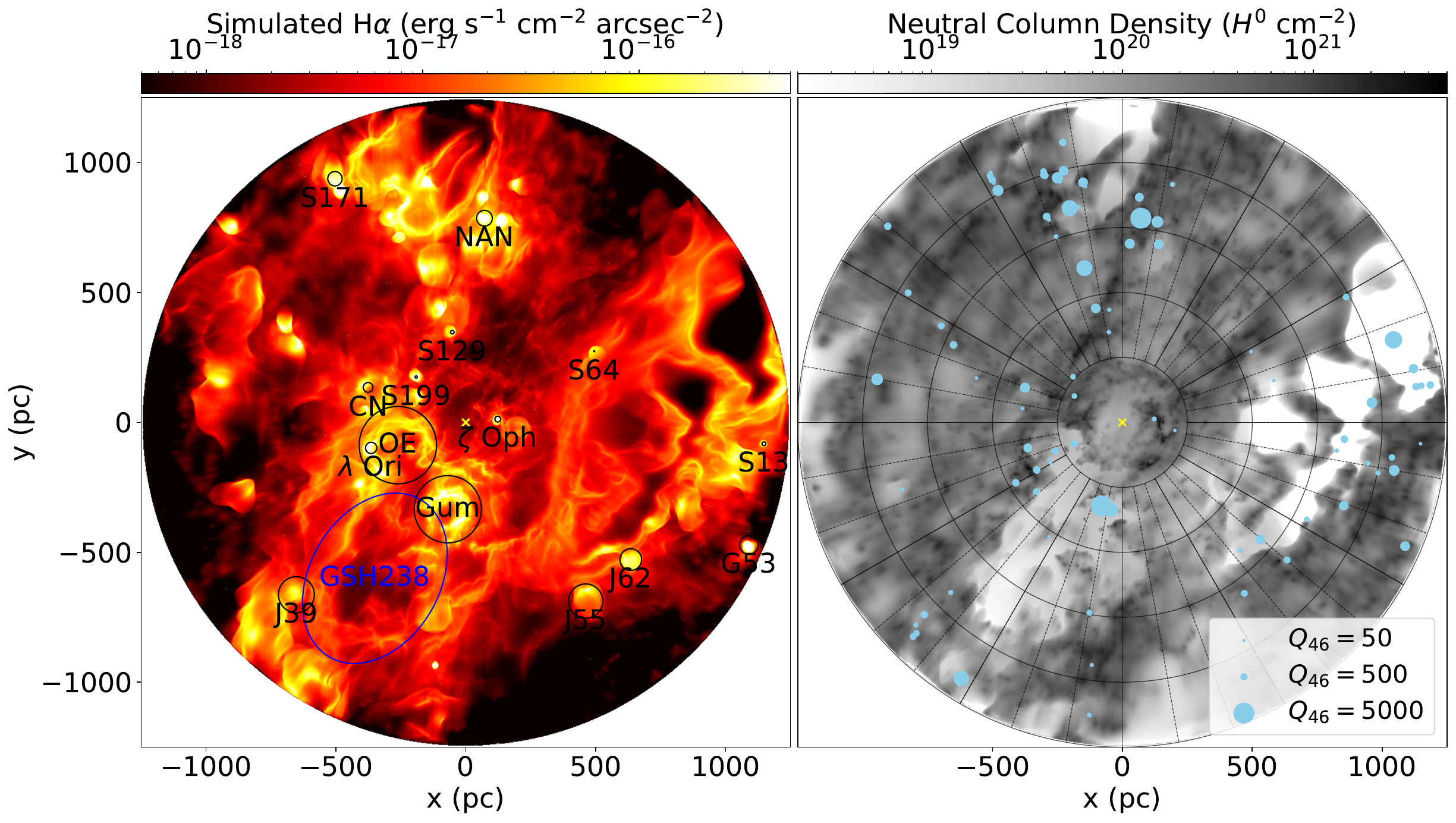}
    \caption{Simulated top down view of the galactic plane around the sun in H$\alpha$ (left) and simulated neutral hydrogen column density (right). The circled regions are the same as in figure~\ref{fig:halphasky}, with the addition of the blue ellipse outlining supershell GSH 238+00+09 (labeled GSH238). The light blue circles on the right panel show our O star locations, with the sizes denoting the ionizing luminosities in $Q_{46}$ ($\rm{s}^{-1}/10^{46}$). The yellow crosses denote the location of the sun in both panels. The galactic center is to the right. In this H$\alpha$ projection it appears that Orion-Eridanus and the Gum nebula are venting into GSH 238+00+09. For a 3D interactive figure, visit \url{https://wim.meikleobney.net}}
    \label{fig:faceon}
\end{figure*}

We find that photons from O stars in the superbubble are in general trapped within the low density void ($n_{H} \approx10^{-3} ~\rm{cm}^{-3}$), and do not ionize through the dense walls ($n_{H} \approx10 ~\rm{cm}^{-3}$). However, many holes and channels in the walls of the superbubble are identifiable and allow a significant amount of radiation to leak and produce nearby diffuse ionized gas. The face-on view in Figure~\ref{fig:faceon} shows that the bubble appears to have broken open on the left side as seen from the Sun ($l=220^{\circ}$), potentially ``venting'' into the Galactic supershell GSH 238+00+09 as predicted by \citet{1998ApJ...498..689H}.  

{\it $\lambda$ Orionis H{\sc ii} region (Sh 2-264)}:  Projected within the Orion-Eridanus complex ($l = -165^\circ$, $b = -12^\circ$), this H{\sc ii} region \citep{2008hsf1.book..757M}  is nearly circular in the WHAM data. As with $\zeta$ Oph, it is present in the simulation, but slightly larger than observed. In 3D this H{\sc ii} region appears as a hemi-spherical outward protrusion in the back wall of the Orion-Eridanus superbubble. The highest density in the dust map along the line of sight to $\lambda$ Ori is at a distance of 420~pc, within one sigma of the distance to $\lambda$ Ori: $386 \pm 60$~pc. Slightly increasing the stellar distance would locate this star in a denser environment and produce a smaller, brighter H{\sc ii} region. One could imagine using the combination of the high spatial resolution of the dust maps combined with H$\alpha$ images to refine the positions of the ionizing sources within this volume.

{\it Gum Nebula}: This nebula \citep{2008hsf2.book...43P}, centered on $(l=260^{\circ},b=-10^{\circ})$, consists of a lower higher-density half (the IRAS Vela Shell) and an upper lower-density ionized cap with a larger expansion velocity. It also contains a significant number of ``bright-rimmed clouds'', dense clouds with H$\alpha$ rims ionized by $\zeta$ Pup and $\gamma^{2}$ Velorum.  This nebula is recognizable in our simulation, but the upper filament has a lower H$\alpha$ surface brightness than in the WHAM map. The simulated H$\alpha$ in the region mostly outlines the boundaries of the dense shell and clouds, recently investigated by Gao et al. (2025, submitted). Like the Ori-Eri superbubble, in the face-on view in Figure~\ref{fig:faceon} this nebula appears to be fractured on the side that abuts the supershell GSH 238+00+09.

{\it North American Nebula (Sh 2-117)}: The O-star in our simulation with the highest ionizing luminosity is the O3.5III 2MASS J20555125+4352246 (refered to as the Bajamar star in \citet{apellaniz16}), with an ionizing luminosity of $5.1\times 10^{49} \rm{photons}~s^{-1}$, representing more than 8\% of the total in our simulation. This source is responsible for ionizing the North American Nebula (NGC 7000 or Sh 2-117), and produces the highest H$\alpha$ brightness anywhere on the simulated sky. This is consistent with WHAM observations where it is brightest region within 1.25 kpc. 

\subsection{Constraints on LyC transport and the Galactic WIM}

Our Monte-Carlo codes allows us to distinguish four different photon termination scenarios: absorbed by dust, absorbed by gas of $n>0.1~\rm{cm}^{-3}$, absorbed by gas of $n<0.1~\rm{cm}^{-3}$, or escape from the simulation. 

We find 15\% of photons are absorbed by dust, 64\% ionize the dense gas, 11\% ionize the low density gas, and 10\% escape the simulation box (in all directions). Not all of the escaping photons will reach the intergalactic medium, as we are missing the absorption from the remainder of the galactic disc. Our estimate that 11\% of the ionizing photon budget from O stars goes to the ionization of the diffuse ionized gas shows good agreement with the observationally motivated estimates of  around 15\% \citep{reynolds84,haffner09}.

To ensure these results are well converged with resolution, a lower resolution run was carried out using a Cartesian grid of $512^{3}$ grid cells. The estimates for Lyman continuum escape fraction, dust losses, dense/diffuse gas contributions and high latitude dust scattering are found to agree within 2\% of the total photon budget going from $1024^{3}$ to $512^{3}$. The morphological match is also retained at lower resolution, although some grid artifacts become apparent in structures which are very local to the Sun.

By volume, 41\% of our simulation grid is ionized (neutral fraction $<0.01$), 51\% neutral (neutral fraction $>0.99$), and 8\% partially ionized. The volume filling factor of diffuse ionized gas (density less than $0.1~\rm{cm}^{-3}$) increases with height above the galactic midplane. The DIG filling factor is  0.3 in the midplane, rising to 0.75 at $z = -500~\rm{pc}$ and 0.5 at $z = +500\rm{pc}$. Additional ionizing sources and time-dependent recombination are required to accurately model the vertical extent of the DIG \citep{mccallum24}.  

\section{Conclusions and Future Work} \label{sec:future}

We have presented our initial simulations of the three-dimensional H$\alpha$ sky, made possible by the dust extinction maps of \citet{edenhofer23}, the parallax distances to nearby O stars \citep{2016A&A...595A...1G}, and the radiative transfer methods developed in \citet{wood99}, \citet{wood04}, \citet{cmacionize}, and \citet{cmi2}. The simulated sky bears a striking resemblance to the observed H$\alpha$ sky  \citep{haffner03,2003ApJS..146..407F}, particularly the substructure seen in the Orion-Eridanus superbubble and the Gum Nebula, although the overall H$\alpha$ emission is less than expected.  Three quarters of the observed O-star ionized H{\sc ii} regions within 1.25 kpc appear in our simulations, with varying levels of morphological agreement. Given the promising agreement with the data, we provided initial estimates for the ultimate fate of Lyman continuum photons emitted in the local Milky Way and the first estimates of the observed H$\alpha$ produced by dust scattering within this new realistic density structure. 

Our simulations provide an impressive validation of the dust extinction maps, and add two important new elements to our evolving three-dimensional understanding of the local ISM: the spatial variation in the local hydrogen and helium ionization fractions and the extreme ultraviolet radiation field. This opens numerous avenues to test, refine, and extend these initial simulations, as discussed below. 

{\it Further tests of the simulations:} Our preliminary assessment of the agreement between observed and simulated classical H{\sc ii} regions can be greatly expanded. We find modeled H{\sc ii} regions are in general larger than observed. Obtaining better agreement with sizes, brightnesses, and morphology of observed regions will involve experimenting with the uncertainties in stellar positions, Lyman continuum production, the sharpness of density interfaces, density filling factors, the gas-to-dust ratio (and variations thereof), and parameters determining the dust scattering and absorption. Our experiments so far give us reason to believe that we can improve constraints on all of these factors by comparing simulations with data. With availability of the 12 underlying realizations of the dust map from \citet{edenhofer23}, this also leaves the possibility of doing a Monte-Carlo uncertainty estimation in the H$\alpha$ sky by running a number of independent simulations which sample the various dust map realizations and O star positions.

The agreement between simulations and observations may be further tested by comparing integrated columns of ionized gas with dispersion measures of pulsars with parallax measurements, e.g. \citet{2020ApJ...897..124O}.  In addition, the detection of pulsar scintillation arcs \citep{2022ApJ...941...34S,ocker24}, which have been used as a probe of foreground ionized ``screens'', may be compared with expected locations of ionized screens created by bubble boundaries in the simulations.  

Although we have compared our simulation to the H$\alpha$ sky maps, the same simulations may be used to predict the expected far ultraviolet continuum from two-photon hydrogen recombination emission, radio free-free emission, radio recombination lines, and low-frequency thermal absorption of radio synchrotron emission from the Galactic disk and halo. This may provide insight, for example, regarding a long-standing puzzle about apparent inconsistency in inferred WIM properties based on H$\alpha$ vs. free-free emission, e.g. \citet{2011ApJ...727...35D}. 

{\it Further refinements of these simulations:} Given the under-production of H$\alpha$ emission over much of the sky, it would be valuable to include the effect of ionizing sources in dust maps that extend to larger distances across the Galactic disk. Although these available maps have lower resolution than our adopted map, with Gaia DR4 on the horizon, the quality and extent of the 3D dust maps is expected to increase. More distant ionizing sources will also contribute to the ionization of the high-altitude gas in the solar neighborhood. Adding information about the vertical extent of the ISM above a kiloparsec would also allow for the measurement of the LyC flux as a function of height, improving our understanding of LyC escape from the Galaxy. 

A second avenue of refinement would be the addition of other sources contributing to the ionization of the WIM, namely B stars hot low-mass evolved stars \citep{2011MNRAS.415.2182F}, including subdwarf OB stars, hot white dwarfs, post asymptotic giant branch (AGB) stars and AGB manqué stars. Gaia-based catalogs for several of these classes of sources are now available, e.g. \citet{2021MNRAS.504.2968P}, \citet{2022A&A...662A..40C}, and \citet{2021MNRAS.508.3877G}, although the Lyman continuum production of these different classes of sources is in general not well constrained.  

While the simulations in this letter focused on the H$\alpha$ sky, the equilibrium ionization state of metals is also tracked in the three-dimensional volume. Synthetic emission line maps can be produced for collisionally excited optical emission lines (e.g. [NII]$\lambda$6584~\AA, [OIII]$\lambda$ 5007~\AA). This capability will be of particular use for interpreting data from the Local Volume Mapper's high-resolution optical spectroscopic survey of the Galaxy \citep{drory24} and may also be compared to observations of the diffuse ionized gas of nearby face-on spiral galaxies. 

{\it Further extensions and ramifications of these simulations:} As the work outlined above improves our understanding of the three-dimensional distribution of the electron density and transport of ionizing radiation, it will become possible to use this improved understanding to examine other phases of the Galaxy. In particular, a reliable map of the three-dimensional electron density may be used to better understand the magnetic field structure of the Galaxy as traced by Faraday rotation of pulsars and extra-galactic sources, e.g. \citet{2022ApJ...940...75D} and Faraday synthesis \citep{2005A&A...441.1217B}.  In addition, one may explore the effect of other sources of diffuse ionization on spectral diagnostics, in particular the X-ray radiation from hot bubbles in the interstellar medium and ionization by low-energy cosmic rays. 

Perhaps the most remarkable result is encapsulated in Figure 3, which shows the face-on view of the local part of the Milky Way disk. The 3D dust maps which have enabled this research have a resolution of roughly 1~pc. When viewed from a distance of 2 Mpc, this resolution would be equivalent to 0.1".  This means that the resolution of our face-on map of the local Galaxy is higher than could be achieved by space-based telescopes for any galaxy except for those within our Local Group. This extraordinarily high-resolution view of the Milky Way,  which can be probed in 3D, will be a valuable resource for understanding the unresolved details in the star- forming disks of other galaxies. 

\section*{Acknowledgements}

LM acknowledges financial support from a UK-STFC PhD studentship. DK acknowledges support from HST-AR-17060 provided by NASA through a grant from the Space Telescope Science Institute, which is operated by the Association of 389 Universities for Research in Astronomy, Inc., under NASA 390 contract NAS5-26555. WHAM and LMH are supported by U.S. National Science Foundation award AST 2009276. We thank Michelangelo Pantaleoni for the helpful discussions relating to our sample of O stars.

KW, RAB, and LMH acknowledge the support and encouragement of Ron Reynolds in their early careers which helped to prepare them for the discoveries presented here. Although Ron sadly passed away in April 2024, we fondly remember him as we continue to explore the mysteries of the warm ionized medium of the Galaxy.

\section*{Data Availability}

3D cubes of electron density and H$\alpha$ emissivity, a 2D sky map of H$\alpha$ (at 15 arcmin resolution) as well as a table of O star properties and positions are available for download at \url{https://doi.org/10.5281/zenodo.15041318}.



\bibliographystyle{mnras}
\bibliography{biblio} 








\bsp	
\label{lastpage}
\end{document}